\documentclass[11pt,twoside]{article}
\usepackage{asp2010}

\resetcounters

\begin{document}

\title{Smoothed Particle Hydrodynamics: Things I wish my mother taught me}
\author{Daniel~J.~Price
\affil{Monash Centre for Astrophysics (MoCA), School of Mathematical Sciences, Monash University, Vic 3800, Australia}}

\begin{abstract}
I discuss the key features of Smoothed Particle Hydrodynamics (SPH) as a numerical method --- in particular the key differences between SPH and more standard grid based approaches --- that are important to the practitioner. These include the exact treatment of advection, the absence of intrinsic dissipation, exact conservation and more subtle properties that arise from its Hamiltonian formulation such as the existence of a minimum energy state for the particles. The implications of each of these are discussed, showing how they can be both advantages and disadvantages.
\end{abstract}

\section{Introduction}
 Smoothed Particle Hydrodynamics (SPH, for recent reviews see e.g. \citealt{monaghan05,springel10,price11}) is a unique numerical method for solving the equations of fluid dynamics. On the surface it is a deceptively simple method --- for example a basic, ``no-frills'' SPH code can be written in a matter of a few days --- but deeper down there are some quite subtle differences in behaviour from (though also similarities with) more standard grid-based techniques. Since the true implications of many of these differences and subtleties have only slowly dawned on me over the course of the last 10 years, I have subtitled this contribution ``Things I wish my mother taught me''\footnote{s/mother/PhD supervisor/}, since they are really things one should be aware of from the beginning when starting to use SPH techniques. They are also the source of many misunderstandings of SPH that creep into the scientific literature and in my view should form the basic starting point by which improvements to the basic techniques may be assessed.

\section{Smoothed Particle Hydrodynamics}
\subsection{The fundamentals}
 SPH starts with the answer to the question: ``How do I compute a density from a distribution of point-mass particles?''. It is clear from the outset that the answer should be i) independent of the absolute position of the particles, depending only on their relative separation, ii) independent of arbitrary rotations of the particle distribution and iii) independent of time or the history of the particles. The density in SPH is computed using a kernel weighted sum,
\begin{equation}
\rho({\bf r}) = \sum_{j=1}^{N_{neigh}} m_{j} W(\vert {\bf r} - {\bf r}_{j} \vert, h),
\label{eq:rhosum}
\end{equation}
where $W$ is the \emph{smoothing kernel}, a function only of relative separation (so as to be independent of translation), with weighting that falls off monotonically according to the radial distance from the measured position $\vert {\bf r} - {\bf r}_{j} \vert / h$ (so as to be spherically symmetric and thus independent of rotation). The simplest example of $W$ is the 3D Gaussian, $W = 1/(\sqrt{\pi} h)^{3} \exp{(-r^{2}/h^{2})}$, where $h$ is the smoothing length.

 The density summation is really the fundamental axiom in SPH, the first real step after the basic idea of discretising fluid quantities onto particles of fixed mass that move with the fluid velocity. Immediately one sees that the resolution of the method will follow mass rather than volume as in grid-based techniques, since each particle carries a fixed mass and so density can only go up due to an increase in the local concentration of particles. Furthermore it is clear that mass will be an exactly conserved quantity since particles cannot lose, gain or diffuse mass. What is less clear but all the more remarkable is that, having written down the density sum, the remainder of the non-dissipative part of the numerical algorithm can be derived \emph{from} it, simply by writing down the Lagrangian
\begin{equation}
L_{sph} = \sum_j m_j \left[ \frac12 v_j^2 - u_j (\rho_j, s_j)\right],
\end{equation}
which is nothing more than the kinetic minus the thermal energy ($u$ is the specific internal energy, an assumed function of density and entropy). Using the Euler-Lagrange equations ($d [\partial L / \partial {\bf v}]/dt = \partial L / \partial {\bf r}$) together with the first law of thermodynamics and the gradient of the density sum, the equations of motion for each particle (say $i$) can be derived self-consistently, giving in their simplest form (neglecting corrections due to variable smoothing lengths)
\begin{equation}
\frac{{\rm d}{\bf v}_{i}}{{\rm d}t} = -\sum_{j} m_{j} \left(\frac{P_{i}}{\rho_{i}^{2}} + \frac{P_{j}}{\rho_{j}^{2}} \right) \nabla W(\vert {\bf r}_{i} - {\bf r}_{j}\vert, h),
\label{eq:dvdt}
\end{equation}
which can straightforwardly be shown to be a discrete form of the Euler equation ${\rm d}{\bf v}/{\rm d}t = -\nabla P /\rho$. 
This basic formulation of the discrete equations contains all the key features of SPH. These are:
\begin{itemize}
\item An exact, time-independent, solution to the continuity equation
\item Advection done perfectly
\item Zero intrinsic dissipation
\item Exact and simultaneous conservation of mass, momentum, angular momentum, energy and entropy
\item A guaranteed minimum energy state for the particles
\item Resolution that follows mass
\end{itemize}
Unpacking the implications of each of these can give us a good appreciation of both advantages and disadvantages of SPH for astrophysical fluid dynamics.

\begin{figure}[htbp] 
   \centering
   \includegraphics[width=0.475\textwidth]{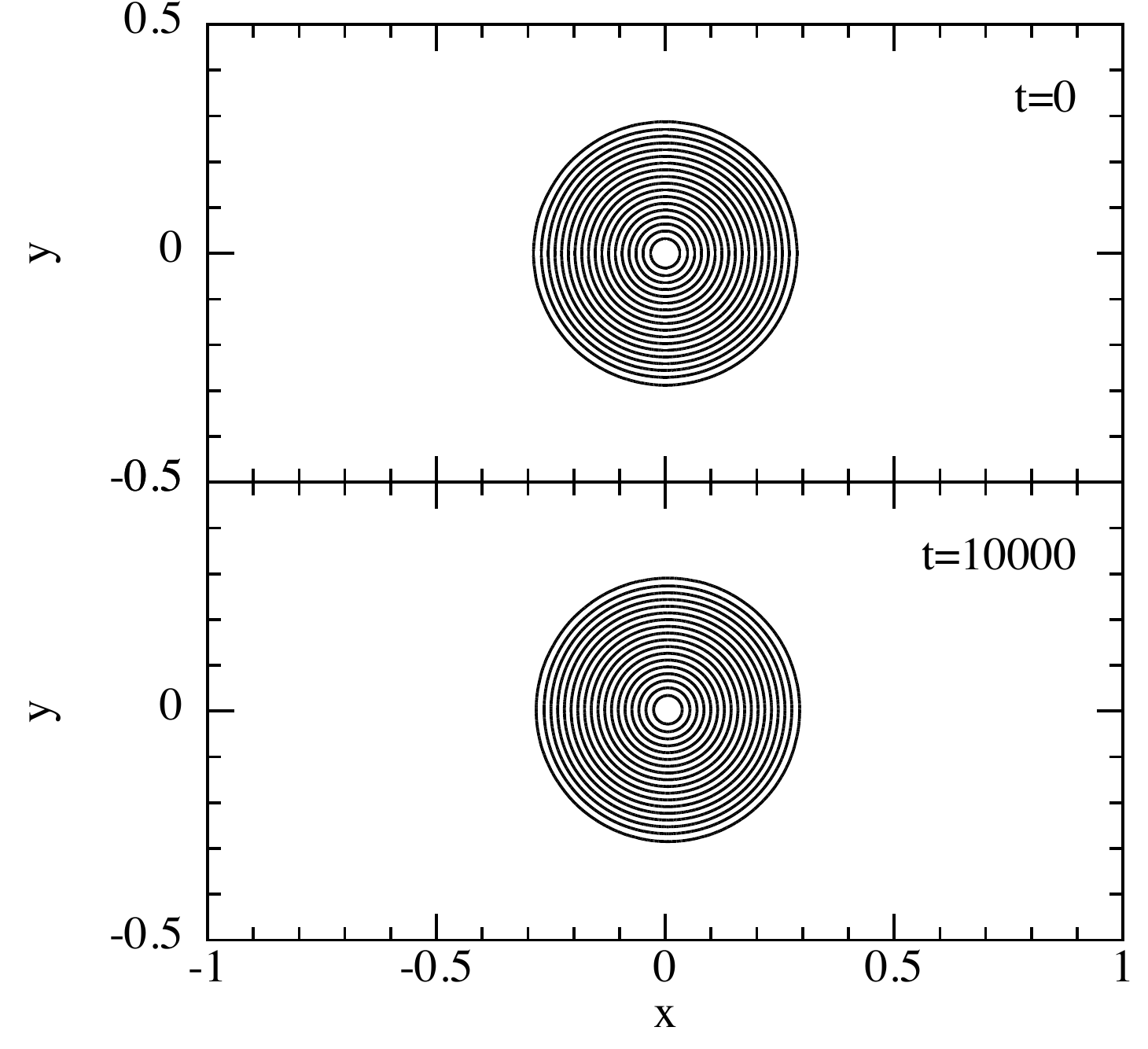} \hspace{0.03\textwidth}
   \includegraphics[width=0.45\textwidth]{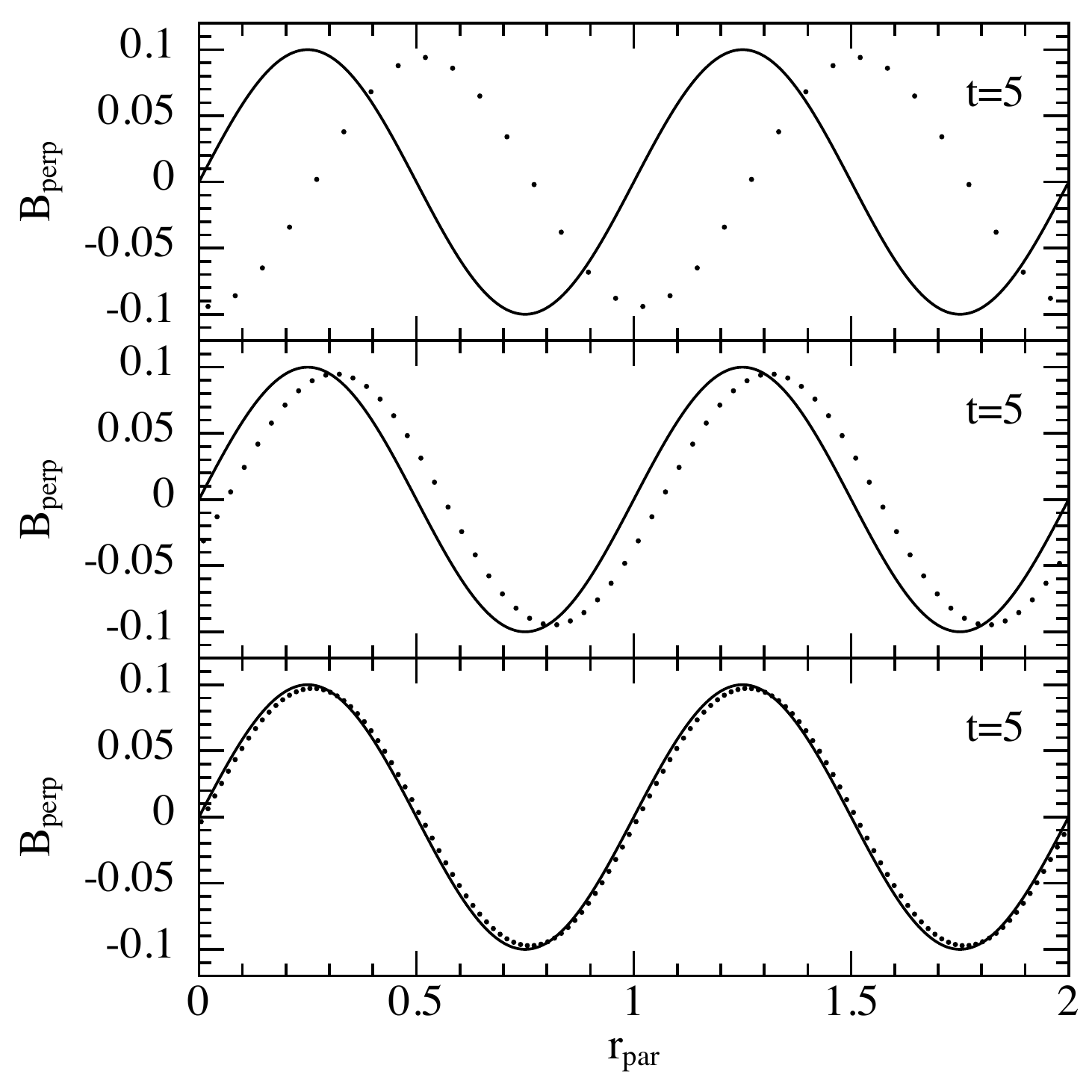} 
   \caption{The meaning of ``exact advection'' and ``zero dissipation'' in SPH. Left: Result of advecting a loop of magnetic current for 10,000 crossings of a 2D domain (bottom panel, showing magnetic field lines), the results of which are identical to the initial conditions (top panel). Right: Evolution of a nonlinear circularly polarized Alfv\'en wave in a 2D domain, showing results after 5 periods at 3 different resolutions (top to bottom). The amplitude is undamped even at the lowest resolution (top panel, $16\times 32$ particles). A small phase error is present that is resolved away at higher resolution ($32 \times 64$ and $64 \times 128$ particles, middle and lower panels).}
   \label{fig:alfvenwave}
\end{figure}

\subsection{Exact advection}
 An example of the exact advection of both mass and other fluid quantities is demonstrated in the left panel of Fig.~\ref{fig:alfvenwave}, showing a loop of magnetic current advected for more than 10,000 crossings of the computational domain. This kind of problem is one of the most difficult tests for Eulerian grid-based codes \citep[see e.g.][]{gs05}, but is essentially trivial in SPH because of the Lagrangian formulation.

\subsection{What ``zero dissipation'' means}
 Perhaps the most misunderstood of the above list is ``zero dissipation''. It should be obvious from the fact that the system is Hamiltonian that the discrete system of particles will be exactly conservative\footnote{At least so far as these properties are also retained in the timestepping algorithm, which we will not discuss further except to note that many so-called `geometric integrators' exist for the symmetry-preserving evolution of Hamilton's equations.} and thus contain zero \emph{intrinsic} numerical dissipation.

\paragraph{Advantages}
 The advantages of having zero dissipation are clear. It means that the numerical solution, in the absence of explicitly added dissipative terms, will neither dissipate nor diffuse energy artificially. An example is shown in the right panel of Fig.~\ref{fig:alfvenwave} which shows the propagation of a non-linear circularly polarised Alfv\'en wave [an exact solution of the equations of magnetohydrodynamics (MHD)], computed with a standard Smoothed Particle Magnetohydrodynamics scheme \citep[e.g.][]{pm05}. The key point is that although there is a phase error at low resolution that is gradually resolved away, the amplitude of the wave is exactly maintained even with a small number of particles since there is no numerical dissipation or diffusion of energy.

\paragraph{Disadvantages}
 The disadvantage of having zero dissipation in the scheme is that, where dissipative terms are required physically, they must be explicitly added. In particular this is the case for shock-capturing, since shocks lead to a physical increase in entropy. Shocks and other kinds of discontinuities are not adequately captured by the Hamiltonian formulation of SPH since in employing the Euler-Lagrange equations we have assumed that the quantities in the Lagrangian (i.e. thermal energy and velocity) are differentiable, implying that discontinuities in those variables need special treatment. Another way to see this is to consider that in arriving at the Euler-Lagrange equations one must neglect certain surface integral terms, corresponding to ``assuming that variations vanish at the boundaries'', which is patently not true when the ``boundaries'' or ``surfaces'' are shocks or other kinds of discontinuities in the physical system.

\subsubsection{Why artificial viscosity is not voodoo magic}
 The addition of an artificial viscosity term to deal with shocks is sometimes regarded as some kind of ``voodoo magic'', and usually the first and only aspect investigated when things don't seem to work (see e.g. the extensive but pointless investigation of artificial viscosity parameters by \citealt{agertzetal}) and blamed as the source for all ills. This should not be the case: Unlike in Eulerian schemes where numerical diffusion is an intrinsic and unavoidable consequence of the discretisation, dissipation terms in SPH are always explicitly added and can thus be directly translated into their physical meaning. Indeed, the standard \citet{monaghan92} or \citet{monaghan97} artificial viscosity terms can be directly translated to a combination of Navier-Stokes bulk and shear viscosity terms, so the parameters and their dependence on numerical resolution are defined analytically and do not particularly need to be ``discovered'' by numerical experiment (despite continuing attempts). In particular, the standard $\alpha$ parameter in the artificial viscosity term provides a shear viscosity $\nu = 1/10 \alpha c_{s} h$ and bulk viscosity $\mu = 5/3 \nu$ in 3D (e.g. \citealt{murray96,lp10}).

\subsubsection{The key is a good switch}
  The first-order dependence of the artificial viscosity parameters on resolution (via the smoothing length, $h$) means that, ironically, SPH has something of a reputation for being \emph{more} dissipative in practice than grid-based codes. Clearly, the solution lies in a good switch which is able to add the correct amount of dissipation at discontinuities while effectively turning off elsewhere, giving back the underlying dissipation-less Hamiltonian SPH formulation. The state-of-the-art in this regard is the viscosity switch proposed by \citet{cd10}, a highly tuned adaptation of the now reasonably well-adopted \citet{mm97} technique. With the same viscosity terms as would be applied to shocks they are able to evolve linear waves for more than 50 periods with essentially no damping of the amplitude. Godunov-SPH schemes \citep[][see also Murante, this volume]{inutsuka02} are another approach to adding dissipation, though the simplest schemes are significantly more dissipative than standard SPH artificial viscosity, while the more advanced schemes add a significant computational overhead without it yet being clear that they are less dissipative overall (e.g. \citealt{cd10} find that their switch fares better, though only on a limited range of test problems).

\begin{figure}[htbp] 
   \centering
   \includegraphics[width=0.495\textwidth]{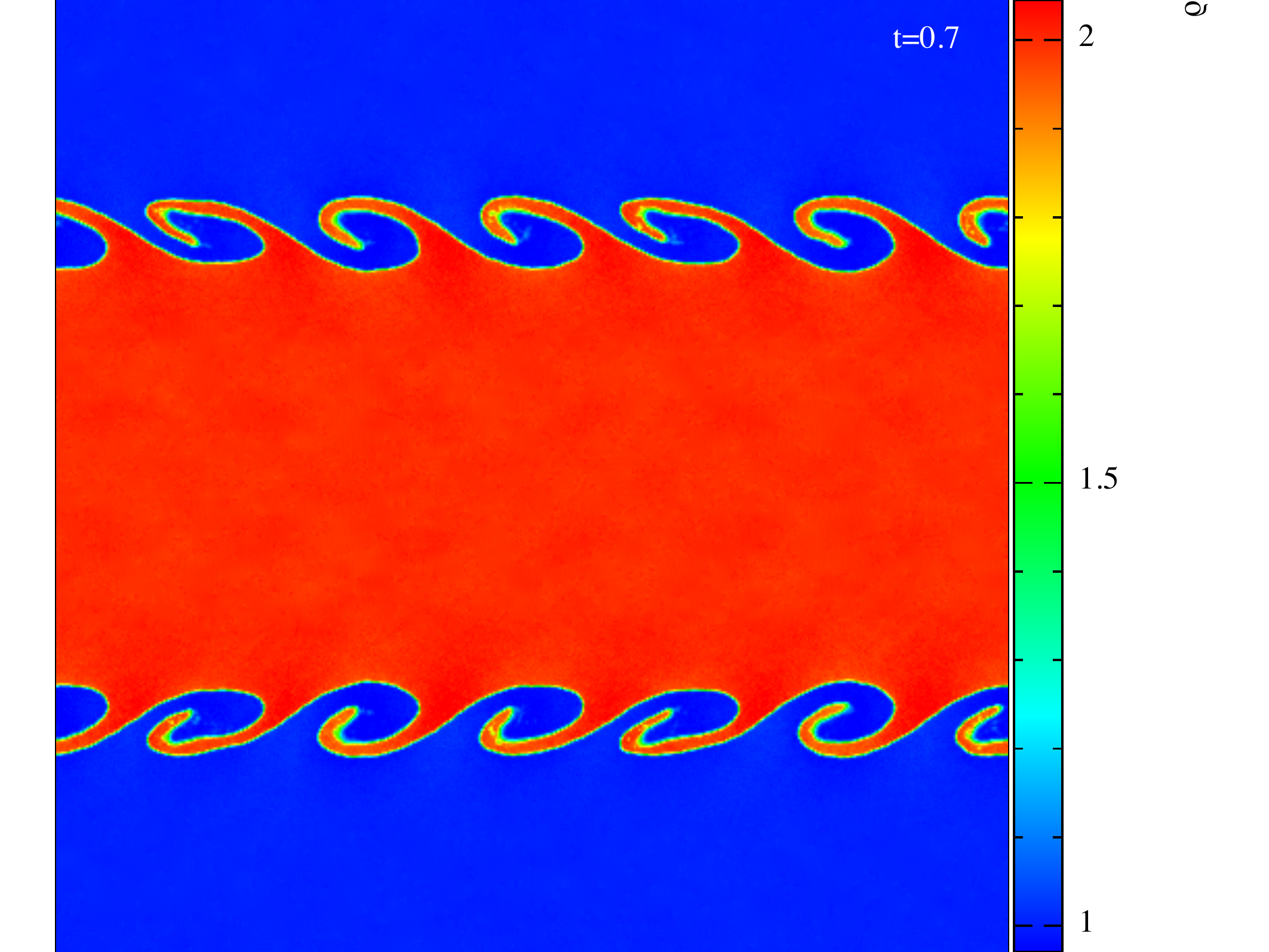} 
   \includegraphics[width=0.495\textwidth]{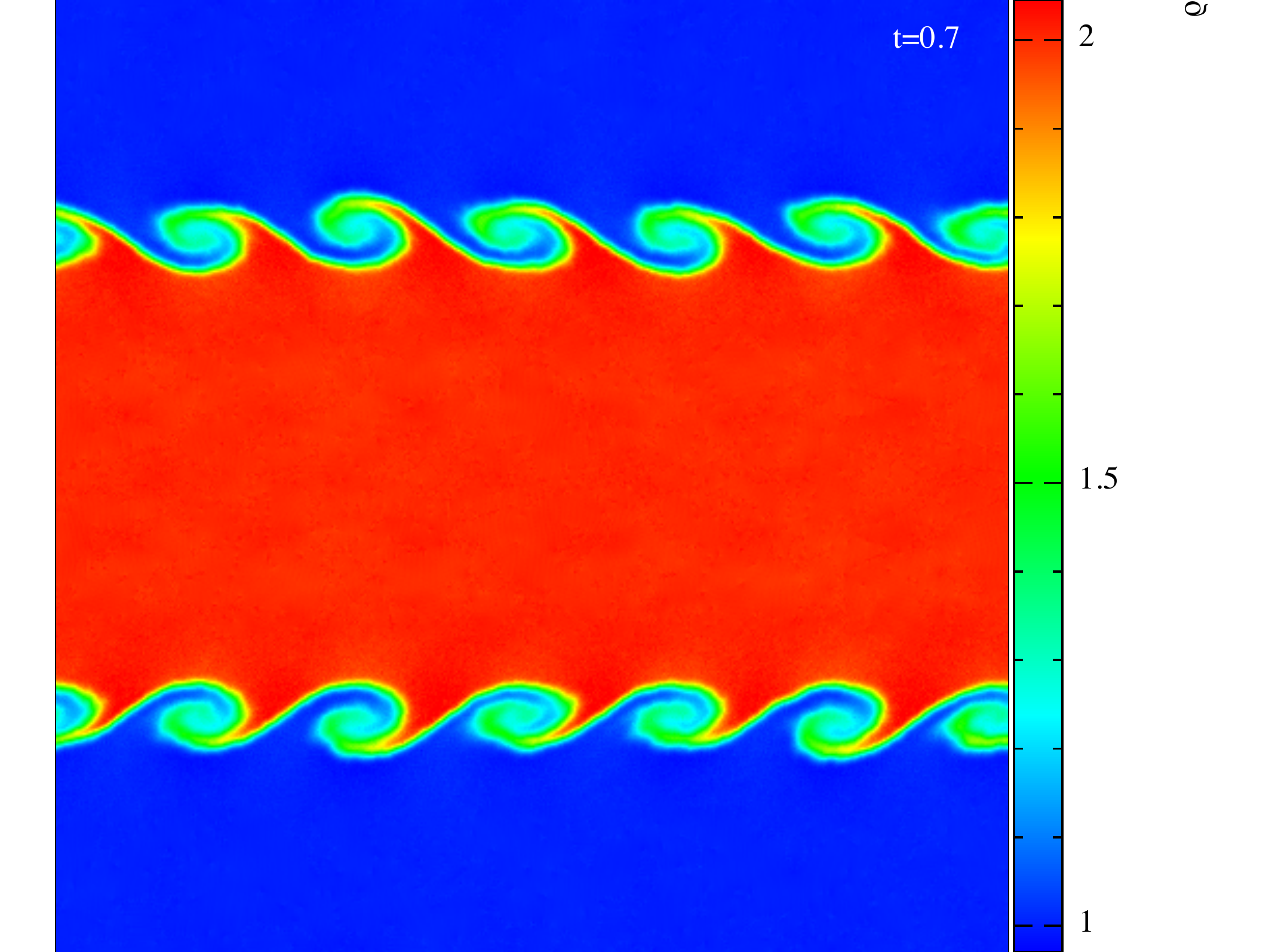} 
   \caption{It is important to ensure that contact discontinuities (discontinuous jumps in density and temperature but not in pressure) are treated correctly in SPH, otherwise mixing can be suppressed across the discontinuity. Left panel shows a Kelvin-Helmholtz instability run across a density ratio of 2:1 with no treatment of the contact discontinuity. Right panel shows same problem with an artificial conductivity term added to treat the discontinuity. Note that the problem has nothing inherently to do with instabilities, but is simply related to treating the discontinuity correctly. See \citet{price08}.}
   \label{fig:kh}
\end{figure}

\subsubsection{The issue with Kelvin-Helmholtz instabilities that has nothing to do with Kelvin-Helmholtz instabilities}
 The need to add dissipative terms to correctly treat discontinuities also explains the issue highlighted in \citet{agertzetal} that has received a great deal of attention in recent years, and has been profoundly misunderstood. For example, it is often asserted --- because the issue in \citet{agertzetal} was demonstrated on a problem involving the Kelvin-Helmholtz (K-H) instability in a shear flow --- that SPH ``has difficulty resolving fluid instabilities''. The issue itself has nothing to do with instabilities, as was already demonstrated in the original \citet{agertzetal} paper where they show that the K-H instability is very well captured if there is no density contrast between the shearing layers, indicating that the problem is more to do with making sure that the density and temperature jump is treated correctly.
 
\begin{figure}[t] 
   \centering
   \includegraphics[width=0.475\textwidth]{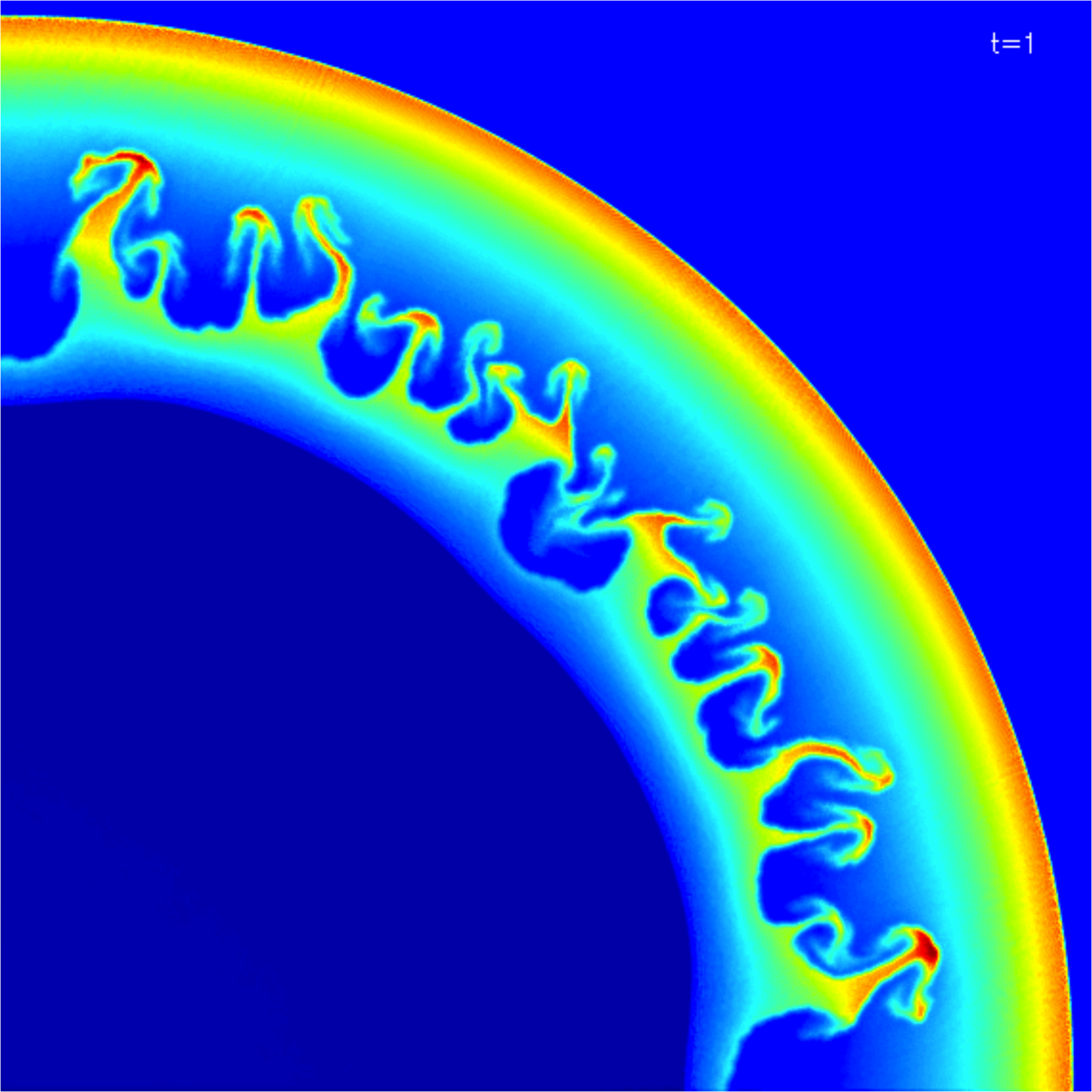} 
   \caption{Example of a Richtmyer-Meshkov instability with SPH, showing density in a 2D blast problem. See \citet{bp09}.}
   \label{fig:instabilities}
\end{figure}

   In \citet{price08} it was shown that the problem is indeed related to the treatment of contact discontinuities (discontinuous jumps in density and temperature where the pressure should remain constant across the jump) in SPH, which in many ``standard'' SPH codes receive no special treatment at all. This leads to a ``blip'' in pressure at the jump which in a 2D or 3D manifests as a suppression of mixing between the shearing layers (see left panel of Fig.~\ref{fig:kh}). With an appropriate treatment of the contact discontinuity, for example using a simple artificial conductivity term proposed in \citet{price08} the blip is removed and the layers mix effectively (right panel of Fig.~\ref{fig:kh}). The more advanced Godunov-SPH schemes (see e.g. Murante, this volume) also perform well on this problem, since they invoke an extra kernel convolution and smoothing of the thermal energy to effectively treat contact discontinuities. Another example is shown in Fig.~\ref{fig:instabilities} showing a Richtmyer-Meshkov instability with a standard SPH + artificial conductivity treatment.

\subsection{Exact conservation}
Since the equations of motion were derived from the Lagrangian using only the density sum, it is clear that the symmetries in the Lagrangian --- i.e., those in the density estimate --- will be reflected as conservation properties in the equations of motion: Galilean covariance, exact conservation of linear momentum (due to the invariance to translations), simultaneous exact conservation of angular momentum (because of the invariance to rotation), energy (because the Hamiltonian is independent of time) and entropy (since the equations were derived under a constant entropy constraint).

\paragraph{Advantages} For astrophysics the simultaneous conservation of linear and angular momentum is crucial. In Eulerian schemes exact angular momentum conservation can only be achieved by using a grid matched to the geometry of the system (e.g. a cylindrical mesh for a cylindrically symmetry accretion disc) and then only when the flow is well aligned with the grid. For SPH the conservation of angular momentum is independent of geometry, meaning that problems involving complicated orbital dynamics are especially well suited to SPH simulation. An example is the study of warped accretion discs by \citet{lp10}, where a detailed verification of predictions from $\alpha$-disc theory was possible because the orbital dynamics is the same regardless of the inclination of the orbital plane and also because the only source of angular momentum dissipation is an explicitly added $\alpha$-viscosity term.

\begin{figure}[t] 
   \centering
   \includegraphics[angle=270,width=\textwidth]{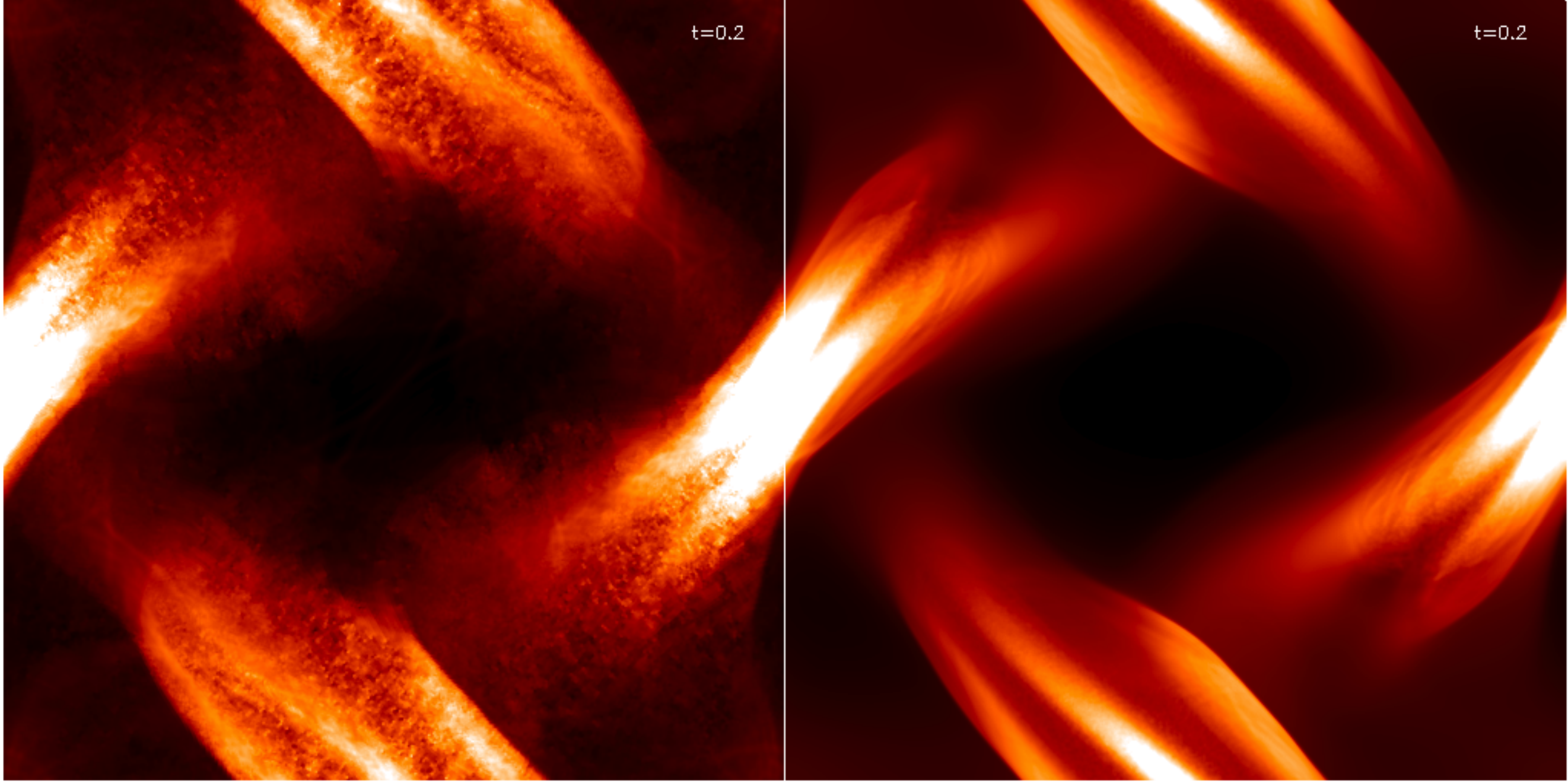} 
   \caption{An example of where exact conservation can be a disadvantage. Left: A snapshot from a calculation of the MHD Orszag-Tang vortex with an unstable formulation of the SPH equations. Right: Same calculation but with a stable formulation. What is remarkable is that energy is better conserved in the Figure on the left. In SPH, noisy particle distributions (e.g. left Figure) rather than code crashes are the main manifestation of a calculation going wrong.}
   \label{fig:orstang}
\end{figure}

\paragraph{Disadvantages} The disadvantages of exact conservation are less obvious, but boil down to the fact that SPH is very robust. Since it can be proven analytically that the dissipation terms always lead to a positive-definite contribution to the entropy \citep[see e.g.][]{pm04a}, it is very difficult to make an SPH code crash! Instead, what tends to happen is that the errors in a calculation are reflected in the particle distribution, which becomes noisy. A particularly extreme example from MHD is shown in Figure~\ref{fig:orstang}, where in this case the more exactly conservative formulation (left panel) is unstable to the tensile (negative-pressure) instability present in conservative stress-tensor formulations of the MHD force, whereas in this case using a non-conservative but stable formulation (right panel) gives much better results.

\subsubsection{How to fix the robustness problem}
 Fixing the robustness problem is fairly straightforward. One simply has to insert a ``code crash'' manually when the particles become noisy. As an aid to the reader, we provide three outlines of such a routine: In C:
\begin{verbatim}
if (particles_are_noisy()) {
   return SPH_HAS_CRASHED;
}
\end{verbatim}
or Fortran
\begin{verbatim}
if (particles_are_noisy()) then
   stop 'sorry, SPH crashed'
endif
\end{verbatim}
and finally, a Python version for inclusion into the AMUSE framework (see Portugeis-Zwart, this proceedings):
\begin{verbatim}
if ( particles ^ AnyofP(``noise'') ):
  die('sorry, your SPH code crashed, we are not AMUSEd')
\end{verbatim}

\begin{figure}[t] 
   \centering
   \includegraphics[width=0.75\textwidth]{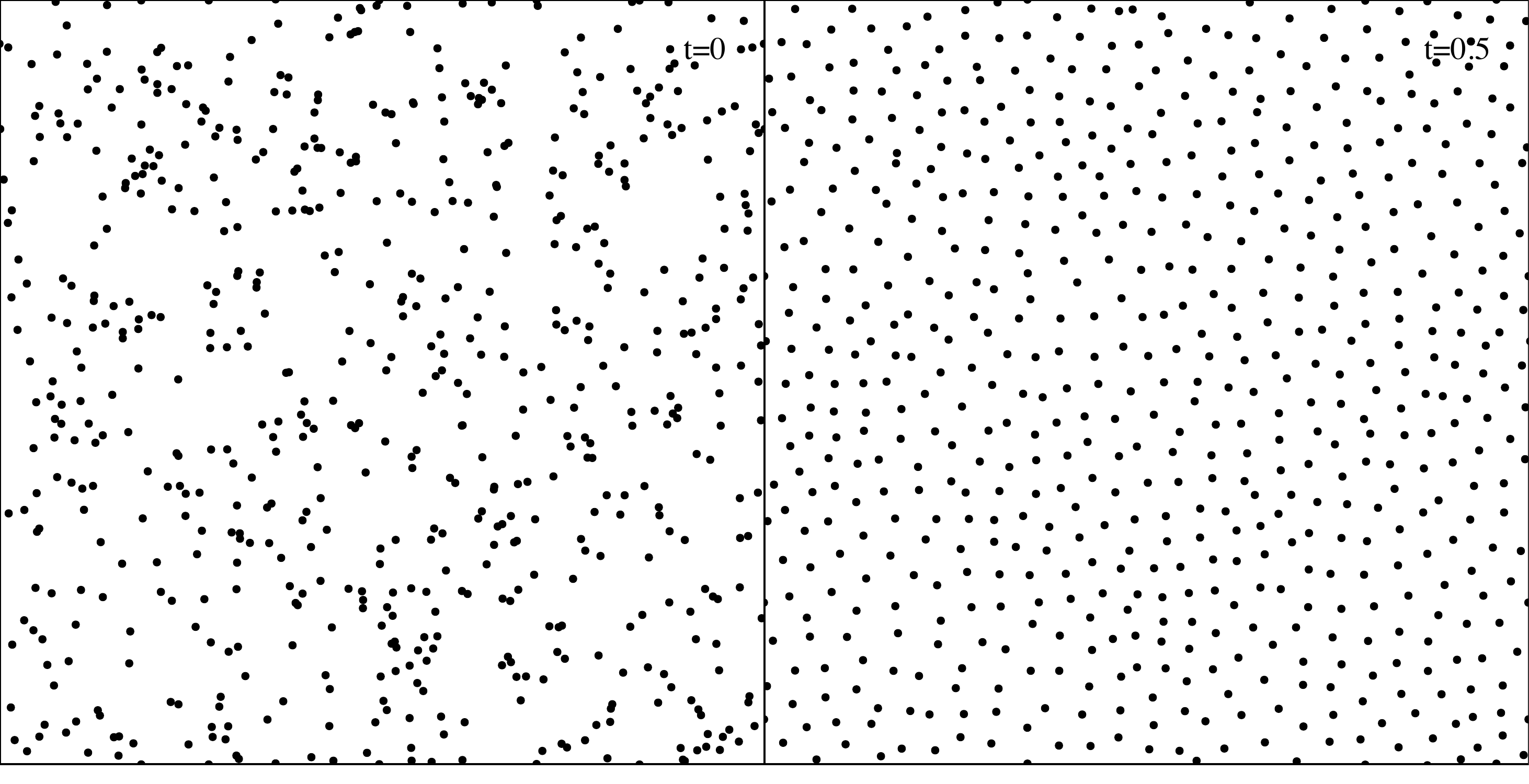} 
   \caption{Relaxation of an initially random particle distribution (left panel, $t=0$) to a glass-like minimum energy state (right panel, $t=0.5$) due to the mutually repulsive force between SPH particles. See \citet{price11}.}
   \label{fig:relax}
\end{figure}

\subsection{The minimum energy state, or the ``grid'' in SPH}
 The Hamiltonian also provides SPH with a very significant advantage over other particle methods by guaranteeing that the particle arrangement will minimise the Lagrangian, implying the existence of a ``minimum energy state'' for the particles. As we have already seen, a stable and noise-free particle arrangement is a very important indicator of the accuracy of SPH solutions, since the arrangement of the particles is similar to the ``grid'' in other Lagrangian methods. 
 
 A good way to understand this is to study what happens to an initially random particle arrangement, e.g. with a constant pressure. With the Hamiltonian formulation (Eq.~\ref{eq:dvdt}) the force does not vanish even though the pressure is constant, leading to a mutual repulsion between particles and thus a rapid adjustment of the random distribution to a ``glass-like'', crystalline state (see Fig.~\ref{fig:relax}). The constraints on the particle distribution imposed by the Hamiltonian nature of the system means that this ``remeshing'' of the particles occurs automatically in SPH, unlike in Lagrangian grid methods where the grid can become arbitrarily distorted. The disadvantage is that this means that there is inevitably some grid-scale motion of the particles which has the potential to overwhelm small perturbations. However, this motion is not in any sense random and is the reason why SPH is able to maintain a good particle arrangement throughout the evolution of quite complicated dynamical processes without the need for explicit grid regularisation or remeshing procedures. It also explains why many alternative particle methods --- those that invoke the exact computation of derivatives instead of conservation --- fail, due to their inevitable production of random, disorganised particle distributions that require complicated re-meshing procedures to fix.

\subsubsection{The tensile instability in SPH}
A disadvantage of using the pressure term to regularise the particle distribution is that it relies on a positive pressure for the particles to be mutually repulsive. For more complicated physical systems such as MHD where the stress can become negative, this can lead to catastrophic instabilities in the particle distribution since particles attract rather than repel each other. This is referred to as the ``tensile instability'' because it also manifests in SPH simulations of elastic dynamics for materials in tension \citep{monaghan00}. In these cases a compromise approach is required where exact conservation is sacrificed slightly in order to gain stability \citep[see e.g.][]{price11}.

\subsubsection{Particle pairing and how to avoid it}
 In SPH it is thus important to check the effects of adjustable parameters (such as the much mis-used ``Number of Neighbours'') on the particle arrangement. In particular, if standard ``Bell-shaped'' or ``Gaussian-like'' smoothing kernels are used for the density estimate (Eq.~\ref{eq:rhosum}), the derivative of these kernels --- as used in the force equation (Eq.~\ref{eq:dvdt}) --- tends to zero at the origin. This means that one cannot simply increase the number of neighbours --- more specifically, the ratio of smoothing length to particle spacing --- arbitrarily for these kernels, since particle pairing occurs as soon as the nearest neighbour is placed sufficiently closely to be ``under the hump'' in the kernel gradient. This means that the force is no longer enough to repel the particles, leading to ``pairing'' or ``merging'' of particles. For the cubic spline kernel (ending at $2h$) pairing occurs when $> 65$ neighbours are employed in 3D (more specifically, when the ratio of smoothing length to particle spacing is $\gtrsim 1.3$) and occurs almost instantly for $h/\Delta > 1.5$ (i.e., $\gtrsim 100$ neighbours in 3D). However, this problem can be easily avoided by simply not using too many neighbours! A better way to reduce kernel bias is to use smoother kernels such as the $M_{5}$ quartic or the $M_{6}$ quintic kernels, which give progressively better approximations to the Gaussian. These kernels employ more neighbours \emph{without changing the ratio of smoothing length to particle spacing} and greatly reduce the ``noise'' associated with particle remeshing \citep[see, e.g.][]{price11}.

\subsection{Resolution follows mass}
 The final key feature of SPH is that resolution is automatically placed where the matter is, a straightforward consequence of having discretised the equations onto Lagrangian particles of fixed mass. For astrophysics this is a strong advantage: It means SPH is very well suited to simulating problems involving gravitational collapse, such as star formation and cosmological structure formation, since computational effort is automatically placed in collapsing, dense objects. The requirement for resolution to follow mass can also be a disadvantage depending on the question being asked. Indeed, the answer to the original subtitle of this paper (SPH: When you should, when you shouldn't) boils mainly down to this.
 
 \begin{figure}[t] 
   \centering
   \includegraphics[width=0.46\textwidth]{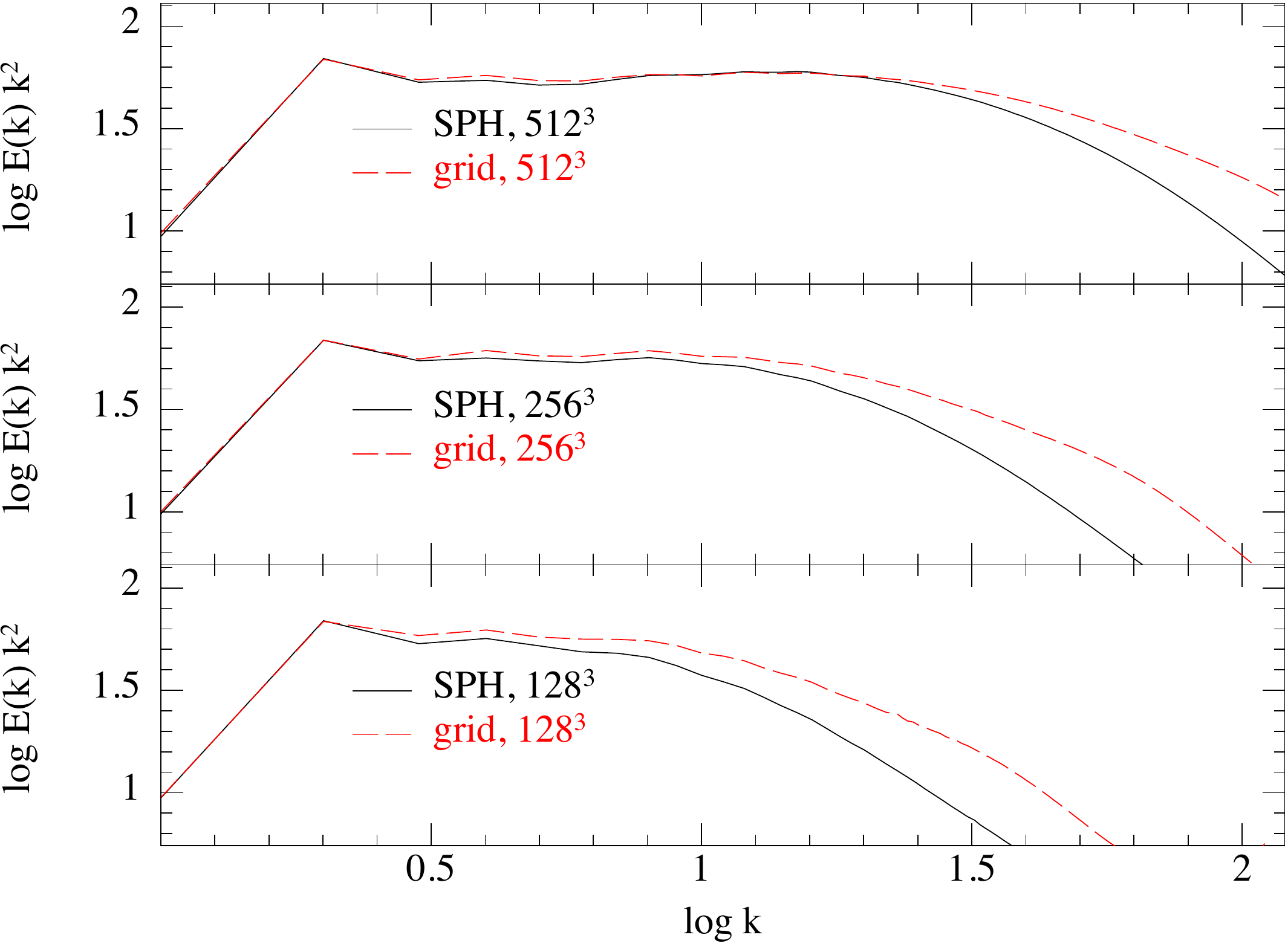} \hspace{0.02\textwidth}
   \includegraphics[width=0.46\textwidth]{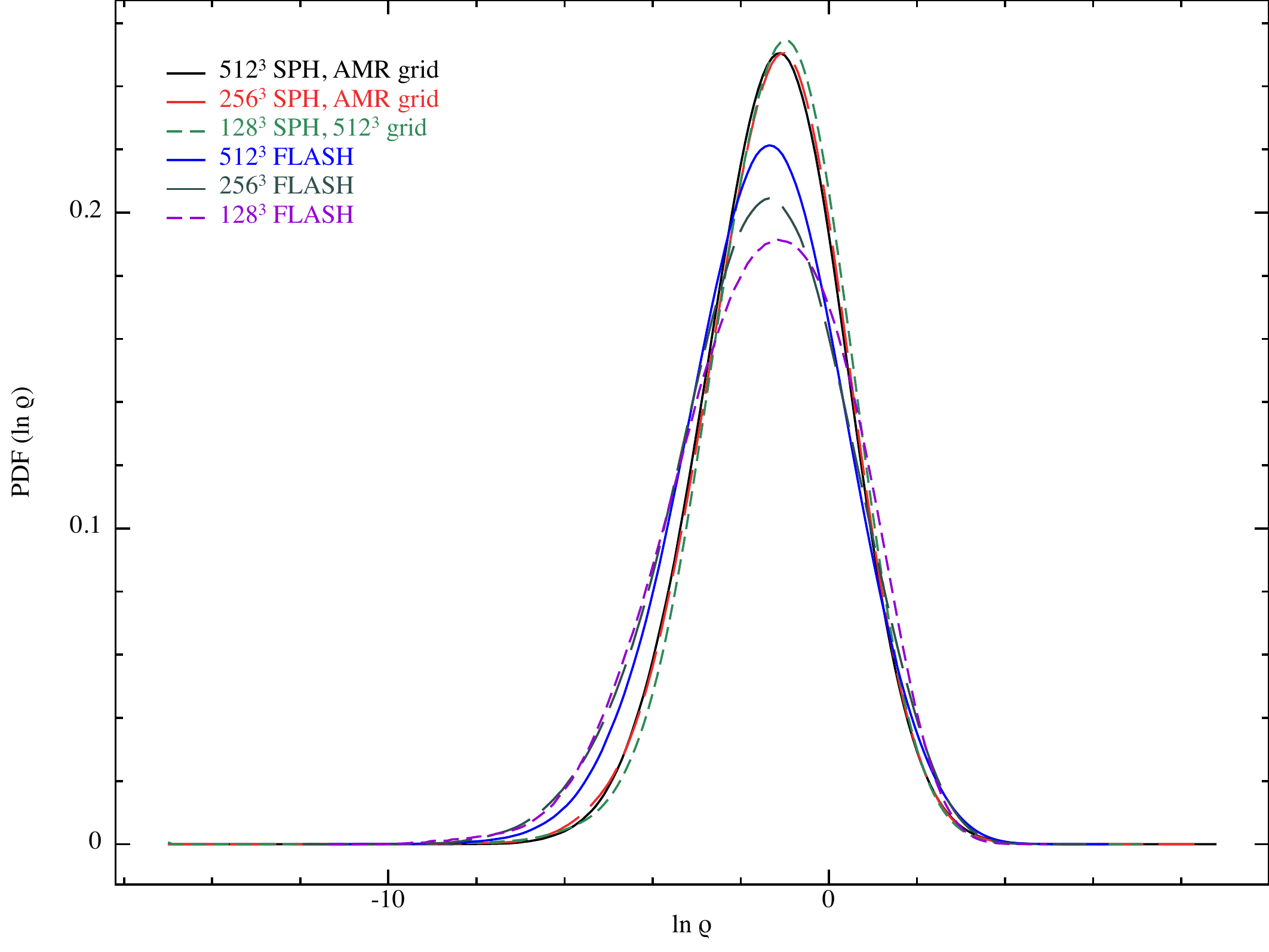} 
   \caption{Kinetic energy power spectra (left panel) and density PDFs (right panel) in an SPH/grid comparison of supersonic turbulence simulations. Since resolution follows mass, SPH was more efficient on questions relating to the density field (right panel), whereas the grid code was more efficient for computing volumetric quantities such as power spectra (left panel). See \citet{pf10}.}
   \label{fig:turbulence}
\end{figure}

  A good example is the recent comparison between SPH and a grid-based code (\textsc{flash}) on simulations of supersonic turbulence in the Interstellar Medium \citep{pf10}. The answer to the basic question of ``Which code should you use?'' depends entirely on which aspect of the problem one desires to study. Fig.~\ref{fig:turbulence} gives two examples. For example, if the question is ``What is the kinetic energy power spectrum?'' (left Figure) then comparable results were obtained only when the number of SPH particles is roughly equal to the number of grid cells, with the grid code thus being the more efficient choice --- by about an order of magnitude in computational cost. However, if the question is ``What is the probability density function of the density field?'' then it was found that SPH resolves densities at $128^{3}$ particles roughly equivalent to the grid code at $512^{3}$ cells --- also reflected in the PDF measurement which was found to be converged at $256^{3}$ particles in the SPH code but not at all converged in the grid code at $512^{3}$ cells (right panel of Figure~\ref{fig:turbulence}). Thus, in this case the SPH code is the more efficient choice --- again by about an order of magnitude in computational cost. However, the main point is that high resolution is required in \emph{both} methods to get similar answers.

\section{Summary}
 I have discussed some of the key features of SPH --- exact advection, zero intrinsic dissipation, exact conservation, the existence of a minimum energy state for the the particles and resolution that follows mass. While these properties have many advantages, they also have disadvantages that need to be understood and accounted for by SPH practitioners. Most importantly, high resolution and resolution studies are a critical requirement with \emph{any} numerical method.

\acknowledgements My thanks go to the conference organisers for a stimulating and very enjoyable conference in Cefal\`u. I used \textsc{splash} \citep{splashpaper}.

\vspace{-0.3cm}
\bibliography{sph,mhd}

\end{document}